\documentclass[a4paper,10pt]{article}
\usepackage[bbgreekl]{mathbbol}
\usepackage[small]{titlesec}
\usepackage[utf8]{inputenc}
\usepackage[T1]{fontenc}  
\usepackage[]{mdframed}
\usepackage[backend=biber,style=alphabetic,natbib=false,giveninits=true,doi=true,isbn=false,url=false,date=year,maxbibnames=99,sorting=ynt,defernumbers=true]{biblatex}
\DeclareNameAlias{default}{family-given}

\DeclareFieldFormat[article]{title}{#1}
\renewbibmacro{in:}{}
\DeclareFieldFormat[article]{pages}{#1}
\DeclareFieldFormat{journal}{#1}
\renewcommand\bf\bfseries

\renewbibmacro*{volume+number+eid}{%
	\printfield{volume}%
	\setunit*{\addnbspace}
	\printfield{number}%
	\setunit{\addcomma\space}%
	\printfield{eid}}
\DeclareFieldFormat[article]{number}{\mkbibparens{#1}}
\DeclareFieldFormat[article]{volume}{\textbf{#1}}
\DeclareFieldFormat{year}{\mkbibparens{#1}}
\DeclareBibliographyDriver{article}{%
	\printnames{author}:%
	\newunit\newblock
	\printfield{title}%
	\newunit\newblock
	\printfield{journaltitle}
	\newunit
	\iffieldundef{number}{\printfield{volume}}{\printfield{volume}\addspace\printfield{number}}
	\addcomma\addspace\printfield{pages}\addspace
	\printfield{year}
}
\AtEveryBibitem{\clearfield{month}}
\AtEveryBibitem{\clearfield{day}}
\usepackage[english]{babel}
\usepackage[T1]{fontenc}
\usepackage{csquotes}
\usepackage{bbm}
\usepackage[leqno]{amsmath}
\usepackage{amsfonts,amsthm,amsbsy,amssymb,dsfont,stmaryrd}
\usepackage[dvipsnames]{xcolor}
\usepackage{braket}

\makeatletter
\newcommand{\leqnomode}{\tagsleft@true\let\veqno\@@leqno}
\newcommand{\reqnomode}{\tagsleft@false\let\veqno\@@eqno}
\makeatother

\usepackage{mathtools}
\usepackage[makeroom]{cancel}

\numberwithin{equation}{section}

\newcommand\myshade{85}
\colorlet{mylinkcolor}{violet}
\colorlet{mycitecolor}{YellowOrange}
\colorlet{myurlcolor}{Aquamarine}

\usepackage[left=2.5cm,right=2.5cm,top=2.5cm,bottom=2.5cm]{geometry}
\usepackage[unicode=true,pdfusetitle,
bookmarks=true,bookmarksnumbered=false,bookmarksopen=false,
breaklinks=false,pdfborder={0 0 1},backref=false,
linkcolor  = mylinkcolor!\myshade!black,
citecolor  = mycitecolor!\myshade!black,
urlcolor   = myurlcolor!\myshade!black,
colorlinks = true,
]
{hyperref}
\usepackage[nameinlink]{cleveref}

\usepackage{tikz}
\usetikzlibrary{arrows}
\usetikzlibrary{intersections}

\usepackage{graphicx}
\usepackage{caption}
\usepackage{slashed}
\definecolor{ct_black}{HTML}{000000}
\definecolor{ct_orange}{HTML}{ED872D}
\definecolor{ct_purple}{HTML}{7A68A6}
\definecolor{ct_blue}{HTML}{348ABD}
\definecolor{ct_turquoise}{HTML}{188487}
\definecolor{ct_red}{HTML}{E32636}
\definecolor{ct_pink}{HTML}{CF4457}
\definecolor{ct_green}{HTML}{467821}

\definecolor{ct2_green}{HTML}{9FF781}
\definecolor{ct2_green_dark}{HTML}{088A08}

\theoremstyle{plain}
\newtheorem{thm}{\protect\theoremname}[section]
\theoremstyle{plain}
\newtheorem{lem}[thm]{\protect\lemmaname}
\theoremstyle{plain}

\theoremstyle{plain}

\theoremstyle{plain}

\theoremstyle{remark}
\newtheorem{rem}[thm]{\protect\remarkname}

\theoremstyle{definition}

\theoremstyle{plain}

\providecommand{\assumptionname}{Assumption}
\providecommand{\claimname}{Claim}
\providecommand{\corollaryname}{Corollary}
\providecommand{\definitionname}{Definition}
\providecommand{\lemmaname}{Lemma}
\providecommand{\propositionname}{Proposition}
\providecommand{\remarkname}{Remark}
\providecommand{\theoremname}{Theorem}
\providecommand{\examplename}{Example}

\crefname{section}{Section}{Sections}
\crefname{appendix}{Appendix}{Appendices}
\crefname{figure}{Figure}{Figures}
\crefname{assumption}{Assumption}{Assumptions}
\crefname{thm}{Theorem}{Theorems}
\crefname{lem}{Lemma}{Lemmas}
\crefformat{equation}{(#2#1#3)}
\crefname{table}{Table}{Tables}

\crefrangelabelformat{equation}{(#3#1#4--#5#2#6)}

\crefmultiformat{equation}{(#2#1#3}{, #2#1#3)}{#2#1#3}{#2#1#3}
\Crefmultiformat{equation}{(#2#1#3}{, #2#1#3)}{#2#1#3}{#2#1#3}

\newtheorem*{lem*}{\protect\lemmaname}

\newcommand{\ZZ}{\mathbb{Z}}

\newcommand{\NN}{\mathbb{N}}

\newcommand{\CC}{\mathbb{C}}

\newcommand{\EE}{\mathbb{E}}

\newcommand{\calH}{\mathcal{H}}

\newcommand\norm[1]{\left\lVert#1\right\rVert}

\newcommand{\tr}{\operatorname{tr}}

\newcommand{\Id}{\mathds{1}}

\usepackage{environ}

\NewEnviron{malign}{%
	\begin{align}\begin{split}
			\BODY
	\end{split}\end{align}
}

\newcommand{\ex}[1]{\EE\left[#1\right]}

\newcommand{\eq}[1]{\begin{align*}#1\end{align*}}
\newcommand{\eql}[1]{\begin{align}#1\end{align}}

\title{Chiral Random Band Matrices at Zero Energy}
\author{\href{mailto:jacobshapiro@princeton.edu}{Jacob Shapiro}\\
	{\footnotesize Department of Mathematics, Princeton University}
}

\begin{document}
	\reqnomode
	
	\maketitle
	\begin{abstract}
		We present a special model of random band matrices where, at zero energy, the famous Fyodorov and Mirlin $\sqrt{N}$-conjecture \cite{fyodorov1991scaling} can be established very simply.
	\end{abstract}
	\section{Introduction}
	Let $W\in\NN$ and $\Set{T_j}_{j=1}^{\infty},\Set{V_j}_{j=1}^{\infty}$ be two sets of indepedent and identically distributed $W\times W$ complex matrices distributed according to the Ginibre ensemble and Gaussian Unitary Ensemble ($\mathrm{GUE}$), respectively.  For any $n\in\NN$ (number of blocks), set $N := nW$ (overall size of the matrix) and define the random $N\times N$ Hermitian matrix 
\eql{
	\label{eq:Hamiltonian}
	H & :=  \begin{bmatrix}V_{1} & -T_{1}^{\ast} & 0\\
		-T_{1} & V_{2} & -T_{2}^{\ast}\\
		0 & -T_{2} & V_{3}\\
		&  &  & \dots\\
		&  &  &  & \dots\\
		&  &  &  &  & V_{n-2} & -T_{n-2}^{\ast} & 0\\
		&  &  &  &  & -T_{n-2} & V_{n-1} & -T_{n-1}^{\ast}\\
		&  &  &  &  & 0 & -T_{n-1} & V_{n}
	\end{bmatrix}\,.
} Here, we use the density of the $\mathrm{GUE}$ or Ginibre ensemble proportional to \eql{\label{eq:density}A\mapsto\exp\left(-W\norm{A}^2_{\mathrm{HS}}\right)} where $\norm{A}^2_{\mathrm{HS}}\equiv\tr\left(|A|^2\right)$ is the Hilbert-Schmidt norm.

	One is interested in the asymptotic behavior of this matrix as both $n,W\to\infty$. To calibrate, when $W\sim O(1),n\to\infty$ one obtains a finite-volume one-dimensional Anderson model (with disordered hopping), which is known to be completely localized whereas for $n\sim O(1)$ one obtains essentially a GUE matrix, which is delocalized. The Fyodorov and Mirlin $\sqrt{N}$-conjecture \cite{fyodorov1991scaling} states that the crossover happens precisely when $W \sim \sqrt{N}$. 
	
	There are many ways to characterize "localized" vs. "delocalized", but in the interest of brevity let us merely phrase one concrete aspect of the problem:
	
	Let $K\subseteq\CC$ be compact. Then, show there exists an $s_0\in(0,1)$ such that for all $s\in(0,s_0)$ and $z\in K$ there exist $C<\infty,\mu>0$, independent of $n$ and $W$, such that
		\eql{\label{eq:fmc}
			\sup_{z\in K}\ex{\norm{\left(H-z\Id\right)^{-1}_{\quad x,y}}^s} \leq W^C \exp\left(-\mu \frac{|x-y|}{W^\sharp}\right)\qquad(x,y\in\Set{1,\dots,n})\,.
		}
	Here by $\left(H-z\Id\right)^{-1}_{\quad x,y}$ we mean the $W\times W$ block matrix element and $x,y$ are the block indices. In the original matrix indices, $i\sim x W,j \sim y W$, the decay is $\exp\left(-\mu|i-j|/W^{\sharp+1}\right)$. For our purposes then, the $\sqrt{N}$-conjecture will be to establish this claim with $\sharp=1$, which corresponds to a scaling $W\sim\sqrt{N}$.
	
	In the pioneering work \cite{Schenker2009} it was first established with $\sharp=7$, and subsequently improved to $\sharp=6$ in \cite{10.1093/imrn/rnx145}. Recently, two contemporaneous preprints appeared \cite{cipolloni2022dynamical,chen2022random} improving the statement to $\sharp=3$. A recent preprint \cite{goldstein2022fluctuations} claims to solve the problem up to (unimportant) logarithmic corrections. We refer the reader to the introduction of \cite{cipolloni2022dynamical} for more context on the problem, different characterizations of localization, and a more complete account of related results.
	
	Here our goal is somewhat perpendicular to the works cited above. We merely wish to remark that if one chooses very special distributions for $\Set{T_j}_{j=1}^{\infty},\Set{V_j}_{j=1}^{\infty}$, and insisted only on studying the resolvent at $z=0$, then \cref{eq:fmc} can \emph{easily} be established with $\sharp=1$. Changing the distributions of the random blocks may seem somewhat benign and should not at all affect the underlying physics. More severe however is the restriction to $z=0$ which means we say nothing about quantum dynamics unfortunately. Be that as it may, it is perhaps worthwhile to make this remark as it offers an extremely simple and immediate way to explain why the critical scaling should be exactly $W\sim\sqrt{N}$.

	\section{Chiral models}
	In the present context, the word "chiral" refers to the chiral (also known as sub-lattice) symmetry which is one of the Altland-Zirnbauer symmetries (in particular it is the composition of time reversal and particle-hole symmetries) \cite{AltlandZirnbauer1997}. More abstractly it may be understood as any self-adjoint unitary operator on a Hilbert space, i.e., a grading operator. According to the convention (which is common in condensed matter physics), a Hamiltonian is considered "chiral symmetric" iff it is odd w.r.t. the chiral grading. Hence, if $\Pi:\calH\to\calH$ is the chiral symmetry operator, $H$ is chiral iff \eql{\label{eq:chiral symmetry}\Set{H,\Pi}=0\,.}
	
	Translating this to the one-dimensional setting in which \cref{eq:Hamiltonian} is embedded, we may take $\calH=\ell^2(\Set{1,\dots,n})\otimes\CC^W$ and $\Pi := (-1)^X\otimes\Id_W$ with $X$ the position operator on $\ell^2(\Set{1,\dots,n})$. This is just a complicated way to say that $\Pi$ is a block-diagonal matrix given by \eq{\Pi = (-\Id_W)\oplus\Id_W\oplus(-\Id_W)\oplus\dots\oplus\Id_W\oplus(-\Id_W)\,.} Then \cref{eq:chiral symmetry} applied on \cref{eq:Hamiltonian} implies that $V_j = 0$ for all $j=1,\dots,n$. The localization of such a random model was studied in \cite{Shapiro2021} (see also \cite{Graf_Shapiro_2018_1D_Chiral_BEC} for its topological properties and \cite{demoor2023footprint} for the density of states at zero energy). It was derived there that under general regularity assumptions on the distribution of the $T_j$'s (which was taken to be alternatingly-distributed), the model is Anderson-localized at all energies except at zero energy where it may fail. Indeed such a failure should happen precisely when the distributions of the $T_j$'s are all identical.
	
	To understand this phenomenon better, let us write down an explicit formula for the Greens function at zero energy (\cite[Prop 7.1, 7.2]{Shapiro2021}):
	\begin{lem}
		Assume that $V_j=0$ in \cref{eq:Hamiltonian} for all $j$. If $n$ is odd then $H$ is not invertible and if $n$ is even, \eql{\label{eq:exact formula for inverse at zero energy}(H^{-1})_{1,n} = -T_1^\circ T_2 T_3^\circ T_4 \dots T_{n-1}^\circ} with $M^\circ \equiv (M^{-1})^{\ast}$.
	\end{lem}
	
	At this stage it seems judicious to associate the fate of localization of this special chiral model at zero energy, and in particular, the answer to the problem \cref{eq:fmc} at $z=0$, with the scaling (in $W$) of the Lyapunov exponents of the random sequence $\Set{T_{2j+1}^\circ T_{2j}}_{j\in\NN}$. Indeed, those are given by \eql{ \gamma_k := \lim_{j\to\infty}\frac{1}{j}\EE\left[\log\left(\sigma_k\left(T_1^\circ T_2 \dots T_{2j-1}^\circ T_{2j}\right)\right)\right]\qquad(k=1,\dots,W)} where $\sigma_k$ is the $k$-th singular value of a matrix (in descending order), and the limit exists if the distributions of the $T_j$'s are sufficiently regular. Then, using similar arguments as in e.g. \cite[(12.87)]{aizenman2015random}, $\gamma_1$ will decide the fate of localization \cref{eq:fmc}, and the $\sqrt{N}$-conjecture becomes \eql{\gamma_1 \sim -\frac{1}{W}\,.}
	
	To calibrate our intuition: it is clear that, e.g., if $W=1$, all matrices commute and the Lyapunov exponent may be readily seen to equal \eq{ \gamma_1 = \EE\left[\log\left(|T_2|\right)\right]-\EE\left[\log\left(|T_1|\right)\right] } which is why if the two distributions are equal, we get a zero exponent, and hence, no exponential decay of the Greens function (i.e. delocalization).
	
	Let us proceed to obtain the scaling as $W\to\infty$. \emph{We make a further crucial simplification to the model}: let us furthermore assume that $T_{2j+1}=\Id$ (instead of being random) and $T_{2j}$ are still chosen from the $\mathrm{Ginibre}(W)$ ensemble. Then, $\gamma_k$ are merely the Lyapunov exponents associated with a sequence of iid $\mathrm{Ginibre}(W)$ matrices. Now we may employ a classical result due to Newman \cite{Newman1986}:
	\begin{thm}[Newman] 
		For an iid sequence of $W\times W$ Ginibre matrices with density \cref{eq:density}, the $k=1,\dots,W$ Lyapunov exponent is given by \eql{\label{eq:Newmans formula} \gamma_k = \log\left(\frac{1}{\sqrt{W}}\right)+\frac12\left(\log\left(2\right)+\Psi\left(\frac{W-k+1}{2}\right)\right)} with $\Psi=\partial\log\circ\Gamma$ the digamma function. 	
	\end{thm}
	
	This explicit formula yields the large $W$ asymptotics as \eq{\gamma_k \sim -\frac{k}{2W}\qquad(k=1,\dots,W)\,.}

	Collecting these observations together we obtain
	\begin{thm}\label{thm:chiral RBM zero energy loc}
		In a "chiral" version of \cref{eq:Hamiltonian} where $n$ is even and where $V_j=0,T_{2j-1}=\Id_W$ for all $j$, and where $T_{2j}$ are all still distributed with $\mathrm{Ginibre}(W)$, \eq{\lim_{n\to\infty}-\frac{1}{n}\log\left(\EE\left[\norm{\left(H^{-1}\right)_{1,n}}\right]\right)\gtrsim\frac{1}{W}} in accordance with the $\sqrt{N}$-conjecture.
	\end{thm}
	A few remarks are in order:
	\begin{rem}
		Apparently for the restricted $z=0$ problem we are studying here it is not necessary to take fractional moments as a regularizing device.
	\end{rem} 
	\begin{rem}
		In deriving \cref{eq:Newmans formula} from \cite[Eq-n (7)]{Newman1986}, we made use of the specific choice of density \cref{eq:density}. This latter choice is conventional and is made to ensure the spectrum of our matrices is of order one, but now it actually has implications on the decay of the Greens function.
	\end{rem}
	\begin{rem}
		Since the classical result of \cite{Newman1986}, there has been continued interest in the topic of Lyapunov exponents of iid sequences of Ginibres. We don't survey these results but mention \cite{Hanin2021} which goes in the direction of performing asymptotics of the singular values of a product of random matrices, and may thus allow one to avoid calculating Lyapunov exponents of an infinite product. We do not pursue this direction here. The particular constraints on $n,W$ in \cite{Hanin2021} do not seem to allow one to explore the delocalized $n\ll W$ regime.
	\end{rem}

		It is tempting to ask whether anything could be said about the full model \cref{eq:Hamiltonian}, given \cref{thm:chiral RBM zero energy loc}. Let us phrase this question in a suggestive way: on $\ell^2(\ZZ)\otimes\CC^W$, let $A$ be a diagonal operator and $B$ be a localized operator. We imagine the particular choice $A=\dots\oplus(V_1-z\Id)\oplus (V_2-z\Id)\dots$ and $B$ the zero energy chiral operator, though the question should be understood abstractly. Can one establish localization for $A+B$? It seems like the answer should be mostly "no" unless $[A,B]=0$ (whence it is trivial). A simple counter example would be one where $B$ is localized at $z=0$ but not throughout its spectrum (for example take the random polymer model \cite{Jitomirskaya2003}) and $A=-z\Id$ is chosen so that $z$ is the special "critical" energy value where $B$ is delocalized. This counter example does not apply in our case since the chiral model is known to be localized at all non-zero energies \cite{Shapiro2021}, although, the issue is we do \emph{not} have good quantitative bounds on the localization length at non-zero energies.

	\bigskip
	\bigskip
	\noindent\textbf{Acknowledgments.} 
	I am indebted to Michael Aizenman, Giorgio Cipolloni, Gian Michele Graf, Ron Peled and Jeffrey Schenker for many stimulating discussions on this special model. I am thankful to Yan Fyodorov for his useful comments and his encouragement to indeed write up this remark. 
	\bigskip


		\begingroup
		\let\itshape\upshape
		\printbibliography
		\endgroup
	\end{document}